\documentclass[12pt]{article}
\usepackage{amsfonts,epsfig}
\usepackage{amsmath,amsfonts}
\usepackage{amssymb,graphicx}
\def\na{\nabla}
\def\pa{\partial}

\def\p{\varphi}

\def\d{\delta}
\def\h{\hat}

\def\t{\tilde}
\def\f{\frac}

\def\p{\varphi}

\def\d{\delta}
\def\h{\hat}

\def\t{\tilde}
\def\f{\frac}

\def\dt{\dot}

\def\l{\label}
\def\e{\varepsilon}
\def\a{\alpha}

\def\la{\lambda}

\def\m{\mu}
\def\n{\nu}
\def\na{\nabla}
\def\r{\rho}
\def\s{\sigma}
\def\S{\Sigma}

\def\th{\theta}
\def\be{\begin{equation}}
\def\ee{\end{equation}}
\def\ba{\begin{eqnarray}}
\def\ea{\end{eqnarray}}

\begin{document}

%%%%%%%%%%%%%%%%%%%%%%%%%%%%%%%%%%%%%%%%%%%%%%%%%%%%%%%%%%
\begin{center}
{\large \bf The notes on thin shells }
 \end{center}
\vspace*{1cm}
\centerline{\bf Mikhail Z. Iofa
\footnote{ e-mail:iofa@theory.sinp.msu.ru}}
\centerline{Skobeltsyn Institute of Nuclear
Physics}
\centerline{Moscow State University}
\centerline{Moscow 119991, Russia}

\begin{abstract}

Geometry of the spacetime with a spherical shell embedded in it is studied 
in two coordinate systems - in Kodama-Schwarzschild coordinates
and in Gaussian normal coordinates. We consider transformations between the
coordinate systems as in the 4D spacetime so as at the surface $\S$  swept in 
the spacetime by the spherical shell. Extrinsic curvatures of the surface swept 
by the shell are calculated in both coordinate systems. 
 Applications to the Israel junction conditions are discussed.

\end{abstract}
%%%%%%%%%%%%%%%%%%%%%%%%%%%%%%%%%%%%%%%%%%%%%%%%%%%%%%%%%%%%%%%%%%%%%%%%%%%%%%%%%%%%%
\section{Introduction}

Thin shells provide a useful tool to study dynamics of collapsing
body eventually forming black hole and Hawking radiation from the black hole
\cite{brout,padm}

 Dynamics of domain walls was studied by Israel \cite{israel}, 
Poisson \cite{poiss},
Ipser and Sikivie \cite{ipser}, Berezin, Kuzmin and Tkachev \cite{berez},
Blau, Guendelman and Guth \cite{blau}, Chowdhury \cite{chowdh},
Gladush \cite{gladush}, Kraus and Wilczek \cite{kraus} and many other authors.

There are two natural settings to study geometry of the  space-time with a
spherical shell: that based on Kodama-Schwarzschild coordinates
and that employing the Gaussian normal coordinate system.

Kodama found that in any (possibly time-dependent) spherically
symmetric space-time there exists a conserved vector which is
timelike in the exterior of the shell. Although the Kodama vector
does not reduce to the Killing vector even in the static space-time,
it can be used to define a preferred "time coordinate" and to
construct a geometrically preferred coordinate system for a
spherically symmetric space-time \cite{kodama,hayw,viss}. Because the Kodama vector is
orthogonal to $dr$, one can construct the 
time coordinate $t$ so that $dr$ is orthogonal to $dt$ \cite{viss}. Using the
Schwarzschild radial coordinate $r$, one arrives at the diagonal,
time dependent spherically symmetric metric which in this 
parametrization in the $(r,t)$ sector has the metric components  
$g_{rr}= (1-2m(t, r)/r)^{-1},\,\,g_{tt}= b^2 (r,t)(1-2m(t, r)/r) $.

The Gaussian normal coordinate system in the neighborhood of the shell \cite{blau,smaller}
is constructed by using a family of non-intersecting geodesics orthogonal
to the surface $\S$ swept by the shell. 
Coordinates of a point outside of the shell are introduced as the geodesic distance
from the point to the shell along the geodesic orthogonal to the surface $\S$
and coordinates of the intersection point of the geodesic with $\S$.

The aim of the present paper is to study connection between two
approaches. We find explicit coordinate  transformation between 
the Kodama -Schwarzschild and  Gaussian normal
coordinate systems. We show that projections of the metrics on
the shell in both cases are identical. In the general case of
time-dependent metrics we calculate external curvature
of the surface swept by the shell in both coordinate systems. 
Applications of the above results to the Israel junction conditions
are discussed.

%%%%%%%%%%%%%%%%%%%%%%%%%%%%%%%%%%%%%%%%%%%%%%%%%%%%%%%%%%%
\section{Kodama-Schwarzschild coordinates}
%%%%%%%%%%%%%%%%%%%%%%%%%%%%%%%%%%%%%%%%%%%%%%%%%%%%%%%%%

 The (2+1) dimensional hypersurface $\S$ swept by  a  spherically symmetric shell
 divides 4D space-time in two regions
$V^\pm$. 
Any spherically symmetric metric in $D=1+3$ spacetime has the general form 
\be 
\l{1.a}
ds^2 =g_{ab} (x) dx^a dx^b = 
{}^{(2)}g_{ij} (x) dx^i dx^j+ r^2 (x)d\Omega^2 . 
\ee 
Here $x^a = (x^i , x^\a )$, where $x^i$ are coordinates in the base space, $\theta, \p$
are coordinates on the spherically symmetric fibers.  
For any
spherically symmetric spacetime it is possible to introduce a vector 
$k^a =\e_\perp^{ab}\na_b r$, (Kodama vector),  which lies in
the radial-temporal plane,  where $\e^{ab}_\perp$ 
$$
\e_\perp^{ab}=\left(
\begin{array}{cc}
\e^{ij}_\perp & 0\\
0&0
\end{array}
\right).
$$
 By construction  Kodama vector $k$ is orthogonal to $\na_a r$. Choosing the time
coordinate $t$ so that $\pa_t\sim k$ \cite{viss}, one  obtains the metric in
the diagonal form because $k$ and $dt$ are orthogonal to $dr$.
 In the parametrization through the time
coordinate $t$ and Schwarzschild radial coordinate $r$ the metric
can be expressed as 
\be 
\l{2.a}
 ds^2_\pm =-b^2_\pm (r,t_\pm )f_\pm (r,t_\pm )dt_\pm^2
 +\f{dr^2}{f_\pm (r,t_\pm )} +r^2 d\Omega^2
,\ee
where $f_\pm (r,t_\pm )=(1-2m_\pm (r,t_\pm )/r )$. In this parametrization $m(r,t)$ is
interpreted as the quasi-local mass.
Note that in this parametrization the metric is diagonal.

Position of the surface is defined by parametric equations $r=R(\tau ), t_\pm
=T_\pm (\tau )$.
The metrics induced on the shell from the regions $V^\pm$ are
$$
ds^2_\pm =-b^2_\pm (R,T_\pm )f_\pm (R,T_\pm )dT_\pm^2
 +\f{dR^2}{f_\pm (R,T_\pm )} +R^2 d\Omega^2.
$$
 From the requirement that the metrics induced on
$\S$ from both regions $V^\pm$ coincide (first Israel condition)
it follows that
$$
-b^2_+ (R,T_+ )f_+ (R,T_+ )\dt{T}_+^2+f^{-1}_+ (R,T_+ )\dt{R}^2=
-b^2_- (R,T_- )f_- (R,T_- )\dt{T}_-^2+f^{-1}_- (R,T_- )\dt{R}^2 .
$$
By choosing $\tau$ as the proper time on the surface, one obtains
\be 
\l{3.a}
 -b^2_\pm (R,T_\pm )f_\pm  (R,T_\pm )\dt{T_\pm }^2+f^{-1}_\pm (R,T_\pm )\dt{R}^2=-1 
\ee 
and projection of the metric on $\S$ is 
\be
\l{4.a} 
ds^2 =-d\tau^2 +R^2 (\tau )d\Omega^2 .
\ee

%%%%%%%%%%%%%%%%%%%%%%%%%%%%%%%%%%%%%%%%%%%%%%%%%%%%
\section{Gaussian normal coordinates}
%%%%%%%%%%%%%%%%%%%%%%%%%%%%%%%%%%%%%%%%%%%%%%%%%%%%%%%

Gaussian normal coordinate system in $4D$ spacetime with a
hypersurface swept by the  
 spherical shell is introduced starting from a
certain coordinate system $\h{x}^\m$ with a metric $\h{g}_{\m\n} (\h{x})$.
The surface $\S$ is parametrized by coordinates $x^i = (\xi, \theta,\p  )$.
 Consider a
neighborhood of $\S$ with a system of geodesics orthogonal to 	$\S$. The neighborhood
is chosen so that the geodesics do not intersect, i.e. any point in the neighborhood
is located on one and only one geodesic. Let us consider a point in the neighborhood
of $\S$ with the geodesic orthogonal to $\S$ which goes through this point. 
The new coordinate system $x^\m$ is
introduced in the following way \cite{blau,smaller}. 
Three coordinates of a point $p$ outside the shell coincide with the coordinates
$x^i$  of the intersection point of the geodesic going through $p$ with $\S$. 
The fourth coordinate 
of the point is equal to the proper geodesic distance along the geodesic 
from the point $p$ to $\S$.  
The proper length along the geodesic is
\be
\l{1.1}
\eta = \int_0^\eta d\eta'\sqrt{\h{g}_{\m\n}(x^\m  )\f{d\h{x}^\m}{d\eta'}\f{d\h{x}^\n}{d\eta'}}
,\ee
 where  $\eta$ is the affine parameter along the geodesic. 
Expression (\ref{1.1}) is invariant under the coordinate transformations with Jacobian
equal to unity, and we can rewrite (\ref{1.1}) through the new coordinates $x^\m$ and
the metric $g_{\m\n}(x)$. Taking the derivative over $\eta$ from both sides of (\ref{1.1})
over $\eta$, one has
\be
\l{n.1}
 g_{\m\n}(x )\f{\pa x^\m (\eta, x^i )}{\pa\eta} \f{\pa x^\n (\eta, x^i )}{\pa\eta}=1,
\ee
or $g_{\eta\eta}(\eta, x^i)=1$.
The tangent vector to the geodesic ${\pa {x}^\m (\eta ,x^i)}/{\pa\eta}|_{\eta =0}$
is orthogonal to  the vector ${\pa x^\n (0, x^ i)}/{\pa x^i}$  in the
tangent plane to  $\S$.
Orthogonality condition of the tangent vector to 
the geodesic to the vector in the tangent plane to $\S$ is
\be
\l{n.2}
g_{\m\n}(x )\f{\pa {x}^\m (\eta ,x^i)}{\pa\eta} 
\f{\pa x^\n (\eta, x^ i)}{\pa x^i}\bigg|_{\eta =0}=0 ,
\ee
or $g_{\eta i}(0, x^i )=0$.
The metrics in $V^\pm$ are
\footnote{Below, to simplify formulas, we omit the subscript $\pm$ everywhere, where
it does not lead to confusion.}
\be
\l{n.3}
ds^2_\pm =d\eta^2 -p^2 (\eta ,\xi  )d\xi^2 + 
2q (\eta ,\xi  )d\tau d\eta +\r^2 (\eta ,\xi ) d\Omega^2 |_\pm
.\ee
Because of the condition (\ref{n.2}), on the surface $\S$
 the interval reduces to
\be
\l{51.a}
ds^2 = -p^2 (0 ,\xi )d\xi^2 +\r^2 (0, \tau) d\Omega^2.
\ee
On the surface $\S$ reparametrization of $\tau$ allows to set $p^2 (0 ,\xi)=1$, which
 is assumed in the following.
It is seen that one can identify $\tau$ with $\xi$ and $R(\tau )$ with $\r (0, \xi )$.
In the following we use the variable $\tau$.

%%%%%%%%%%%%%%%%%%%%%%%%%%%%%%%%%%%%%%%%%%%%%%%%%%%%%%%%%%%%%%%%%
\section{Transformation between the coordinate systems}
%%%%%%%%%%%%%%%%%%%%%%%%%%%%%%%%%%%%%%%%%%%%%%%%%%%%%%%%%%%%%%%%

Coordinate transformation $t_\pm=t_\pm(\eta ,\tau ),\,\,\, r_\pm =r(\eta
,\tau )$
 from Kodama -Schwarzschild coordinates $ x^\m=(t,r,\theta,\p )$
to Gaussian normal coordinates $\h{x}^\m =(\eta, \tau,\theta,\p )$  yields the following relations
between the components of the metrics (\ref{3.a}) and (\ref{n.3})
 \ba 
\l{7.a}
-b^2 f \dt{t}^2+f^{-1}\dt{r}^2 =-p^2 \\
\l{8.a}
-b^2 f {t'}^2+f^{-1}{r'}^2 =1 \\
\l{9.a}
-b^2 f \dt{t}t' +f^{-1}\dt{r}r' =q,
\ea
where prime and dot denote derivatives over $\eta$ and $\tau$.
On the surface $\S$ transformations (\ref{7.a})-(\ref{9.a}) are of the same form with the
substitution $\dt{t}\rightarrow \dt{T},\,\,\, \dt{r}\rightarrow
\dt{R}$ and $q=0, \,\,\, p = 1 $. 

 It is straightforward to obtain solution of the system (\ref{7.a})-(\ref{9.a}) 
in the spacetime regions $V^\pm$
as $\dot{t}= \dot{t} (p,q,\dot{r}),\,t' = t'(p,q,\dot{r}),\, r' = r'(p,q,\dot{r})$. 
Instead of writing this cumbersome and	not instructive general solution, we
consider the restriction of the transformation to the surface $\S$ which we use below
\ba
\l{13.a}
\dot{t}^2|_\S = \dot{T}^2 = \f{f(T,R)+\dot{R}^2}{b^2 (T,R) f^2 (T,R)},\qquad
{t'}^2|_\S =\f{\dot{R}^2}{b^2 (T,R) f^2 (T,R)}, \qquad {r'}^2|_\S = f(T,R)+\dot{R}^2 . 
\ea
Because $\S$ is orientable, on $\S$ can be defined a normal vector.
In Kodama-Schwarzschild
coordinates tangent vector, $u_\m$, and normal vector, $n_\m$,  at
either side of the surface $\S$ are 
\ba
\l{42.a}
u^\m_\pm =(\dt{T}_\pm , \dt{R}, 0,0),\qquad u^2 =1,\\
\l{52.a}
n_{\m \pm}=\pm N(-\dt{R},\dt{T}_\pm ,0,0), \qquad u^\m n_\m =0, \\
\l{62.a}
n^\m_\pm =\pm N \left(\f{\dot{R}}{b^2_\pm f_\pm},\dot{T}_\pm f_\pm,0,0 \right).
\ea
Normalizing $n^2$ to unity, we obtain
\be
\l{n.4}
n^2 =N^2 \left(-\f{\dot{R}^2}{b^2 f} +\dt{T}^2 f\right)=  \f{N^2}{b^2}=1.
\ee
Transformations of the components of the tangent vector from Kodama-Schwarzschild  
coordinates $(t,\,r)$
to Gaussian coordinates $(\tau ,\eta )$ are
\ba
\l{n.41}
\h{u}^\pm _\eta=u_{\m}^ \pm\f{\pa x^\m_\pm}{\pa\eta}\bigg|_\S=
 -b^2 f\dt{T}t' +\dt{R}r' /f\big|^\pm_\S =0,\\
\l{n.42}
\h{u}^\pm_\tau\bigg|_\S=u_\m ^\pm\f{\pa x^\m_\pm}{\pa\tau}\bigg|_\S =-b^2 f\dt{T}\dt{t}
+\dt{R}\dt{r} /f \big|^\pm_\S =- 1.
\ea
The corresponding transformations of the components of the normal vector are
\ba
\l{n.43}
&{}&\h{n}^\pm_\tau=n_\m ^\pm\f{\pa x^\m_\pm}{\pa\tau}\bigg|_\S=
-\dt{R}\dt{t}+\dt{T}\dt{r}|_\S=0,\\
\l{n.44}
&{}&\h{n}^\pm_\eta=n_\m^\pm\f{\pa x^\m_\pm}{\pa\eta}\bigg|_\S =
N(-\dt{R}{t'}+\dt{T}{r'})|_\S =\pm 1
\l{18.a}.
\ea
In (\ref{n.43})-(\ref{n.44}) we used the expressions (\ref{13.a}), where all the square roots
for $\dot{T}, \,t'|_\S$ and $r'_\S $ are taken with the same signs. The upper sign in (\ref{n.44})
corresponds to the square roots taken with the sign (+).

Next, we consider another method to construct the  explicit form of the coordinate transformation
 from Kodama-Schwarzschild
coordinates to Gaussian normal coordinates. The problem can be solved in principle by solving
 the geodesic equations
\be
\f{d^2 x^\m}{d\la^2}+\Gamma^\m_{\n\la}\f{dx^\n}{d\la}\f{dx^\la}{d\la}=0.
\ee
In the
general case with the metric  (\ref{2.a}) depending on $t$ 
the system of nonlinear differential equations is not
tractable.
Explicit relations can be obtained in the case of the
metric (\ref{2.a}) with the components independent of $t$
\footnote{In this case it is possible to set $b(r)=1$: introducing
new variable $\r$ by the relation $d\r=dr/b(r)$ and denoting $k(\r)= 
b^2 f (r(\r ))$, we obtain the metric (\ref{2.a}) with $b=1$.}.
In this case the geodesic equations in Kodama -Schwarzschild coordinates
are
\ba
\l{d.1}
&{}&\f{d^2 t}{d\la^2} + \f{f_{,r}}{f}\f{dt}{d\la}\f{dr}{d\la}=0 \\\nonumber
&{}&\f{d^2 r}{d\la^2} +\f{f_{,r} f}{2}\left(\f{dt}{d\la}\right)^2 -\f{f_{,r}}{2f}
\left(\f{dr}{d\la}\right)^2 +fr \left(\f{d\theta}{d\la}\right)^2
+ fr\sin^2 \theta\left(\f{d\p}{d\la}\right)^2 =0\\\nonumber 
&{}&\f{d^2 \theta}{d\la^2}=0 = \f{d^2 \p}{d\la^2}
\ea
where $\la$ is affin parameter.
We look for a solution of the form $r=r(\la,\tau), t=t(\la,\tau)$, where dependence
on $\tau$ appears from the boundary conditions on $\S$, and $d\theta/d\la=
d\p/d\la=0$. 
The first integrals of the system of equations are
\ba
\l{13.b}
&{}&\f{dt}{d\la}=\f{C_t (\tau)}{f(r)},\\
\l{14.b}
&{}&\f{dr}{d\la}=\pm\left(f(r)+ C^2_t (\tau )\right)^{1/2}  \\\nonumber
&{}&\f{d\theta}{d\la} = \f{d\p}{d\la}=0.
\ea
By construction the vector $l^\m (\la,\tau)= (\pa t /\pa\la, \pa r/\pa\la, 0, 0)$ is 
 tangent to the geodesic. 
Let us consider the geodesics orthogonal to $\S$. In this
case the affine parameter $\la$ can be identified with 
the parameter $\eta$. From (\ref{13.b})-(\ref{14.b}) it follows that the vector $l_\m$
is normalized to unity.
At the surface $\S$ the vector $l^\m$ up to the sign coincides with the normal vector (\ref{62.a}).
Thus, at the surface  $\S$ we have ${C_t}^2 =\dot{R}^2$. For $l^\m (\la,\tau)$
we obtain
\be 
\l{15.b} 
l^\m   =\pm\left( \f{\dot{R}}{ f(r)}, \sqrt{f(r) +\dot{R}^2},0,0 \right) .
\ee
On the surface $\S$ solution (\ref{15.b})
coincides with the formulas (\ref{13.a}) with no dependence on $T$.
 
In Kodama-Schwarzschild parametrization the variables $\tau$ and $\eta$ 
have a clear geometrical meaning: at the $(t,\,r)$ plane
$\tau$ varies along the trajectory of the shell  $(R(\tau ), \,T(\tau))$,
and $\eta$ varies along the geodesics orthogonal to
the surface swept by the shell.

%%%%%%%%%%%%%%%%%%%%%%%%%%%%%%%%%%%%%%%%%%%%%%%%%%%%%%%%%%%%%%%%%%%%%%%%%%%%%
\section{Extrinsic curvature}
%%%%%%%%%%%%%%%%%%%%%%%%%%%%%%%%%%%%%%%%%%%%%%%%%%%%%%%%%%%%%%%%%%%%%%%%%%%%%%

The extrinsic curvatures at either side of $\S$ are
$$
K_{ij}^\pm=\left(h_\m^\la n_{\n ;\la }\f{dx^\m}{dx^i}\f{dx^\n}{dx^j}\right)^\pm_\S ,
$$
where $x^i =(\tau, \theta,\p)$ are coordinates on $\S$, $h_{\m\n}=g_{\m\n}-n_\m n_\n$,
  and (;) denote covariant derivative
with respect to $g_{\m\n}^\pm $. In Kodama -Schwarzschild parametrization 
$x^\m|_\S =(T(\tau ),R(\tau ),\theta,\p)$.
In Kodama-Schwarzschild parametrization the non-zero components of the 
extrinsic curvature are 
\ba
\l{3.b}
&{}&{K}^{\pm}_{\tau \tau}=\left( h_\m^\la n_{\n ;\la }u^\m u^\n\right)^\pm
=(n_{\n ;\la}u^\la u^\n\,)^\pm ,\\
\l{4.b}
&{}&K^\pm_{\theta \theta }=h_\m^\la n_{\n;\la}\f{dx^\m}{d\theta}\f{dx^\n}{d\theta}
=h^\la_{\theta} n_{\theta;\la}= n^\pm_{\theta ;\theta}=-\Gamma_{\theta \theta}^R n_R |^\pm,\\
&{}&K^\pm_{\p\p}= n^\pm_{\p ;\p} =-\Gamma_{\p \p}^R n_R |^\pm.
\ea
Using the identity $n_{\m;\n}u^\m u^\n = - n_\m u^\m_{;\n}u^\n $,
 we have $K_{\tau\tau}=-u^\n u^\m_{;\n}n_\m$.
From the identity  $u_\n u^\n =1$ it follows that $u^\m u_{\m;\n}u^\n =0$ or
$u_T u^{T}_{;\n}u^\n + u_R u^{R}_{;\n}u^\n =0$.
Direct calculation yields
$$
u^R_{;\n}u^\n =\ddot{R}+\dot{T}^2\Gamma^{R}_{TT}
+\dot{R}^2\Gamma^{R}_{RR}+2\dot{T}\dot{R}\Gamma^{R}_{RT} 
.$$
  $K_{\tau\tau}$ is expressed as
\be
\l{71.b}
K_{\tau\tau}= - u^\n u^\m_{;\n}n_\m = u^R_{;\n}u^\n \left(n_T \f{u_R}{u_T} -n_R \right)
=-\f{u^R_{;\n}u^\n}{\dot{T}b f(R,T)}
.\ee
Using the formulas of Sect.4 for $u_\m$ and $n_\m$, we obtain
the extrinsic curvature in Kodama-Schwarzschild coordinates
\ba 
\l{7.b} 
&{}&
K^\pm_{\tau\tau}
 =-\left(\ddot{R} +\f{f_{,r}(R,T)}{2}+
(f(R,T)+\dot{R}^2 ) \f{b_{,r} (R,T)}{b (R,T)}-
\dot{R}\dot{T}\f{f_{,t}(R,T)}{f(R,T)}\right)\bigg/f b \dot{T}\bigg|^\pm,\\
&{}&K^\pm_{\theta \theta } = R\sqrt{\dot{R}^2 +f(R,T)}\bigg|^\pm , \\
\l{8.b} 
&{}&K^\pm_{\p\p}=R\sqrt{\dot{R}^2 +f(R,T)}\sin^2 \theta\bigg|^\pm  
\l{9.b} 
,\ea
where $\dot{T}= (f(R,T)+\dot{R}^2 )^{1/2}/b(R,T)f(R,T)$.
 
In Gaussian normal coordinates the components of the extrinsic curvature are
\ba
\l{10.b}
\l{11.b}
&{}&\h{K}^\pm_{\tau\tau}=\h{n}_{\tau ;\tau} =
-\Gamma^\eta_{\tau\tau}\h{n}_\eta|^\pm_\S =
-\f{1}{2}(p^{2})_{,\eta}(\tau ,\eta)\h{n}_\eta|^\pm_{\eta=0},\\
\l{12.b}
&{}&\h{K}^\pm_{\theta \theta }= \h{n}_{\theta ;\theta }=
-\Gamma^\eta_{\theta \theta }\h{n}_\eta|^\pm_{\eta=0}
=\f{1}{2}r^2_{,\eta}\h{n}_\eta|^\pm_{\eta=0},
\\\nonumber
&{}&\h{K}^\pm_{\p \p }=\h{n}^\pm_{\p ;\p }=
-\Gamma^\eta_{\p \p }\h{n}_\eta|^\pm_{\eta=0}=
\f{1}{2}r^2_{,\eta}\h{n}_\eta\sin^2\theta|^\pm_{\eta=0}.
\ea
In the general case of the functions $f(t,r)$ depending on $t$ calculation
is straightforward, but cumbersome. Below we perform calculation
for the case of $f^\pm (r)$ independent of $t$.
Using solutions  (\ref{15.b}), 
$$
t'=\f{\dot{R}}{f(r)},\qquad r' =\sqrt{f(r)+\dot{R}^2}
,$$
we have
\be
\l{1.e}
\dot{t}'|_\S =\f{\ddot{R}f(R)-f_{,r} (R)\dot{R}^2}{f^2 (R)}, \qquad
\dot{r}'|_\S = \f{ f_{,r} (R)\dot{R} +2\dot{R}\ddot{R}}{2\sqrt{f(R)+\dot{R}^2}}.
\ee
Using (\ref{7.a}), we obtain
\be
\l{3.e}
 p^2_{,\eta}\big|_\S= (f\dot{t}^2 -f^{-1}\dot{r}^2)_{;\eta}|_\S =
\left[ f_{,r}(R)r' \dot{T}^2 +2f(R)\dot{T}\dot{t}' 
+\f{f_{,r}(R)r'\dot{R}^2}{f^2 (R)}-
\f{2\dot{R}\dot{r}'}{f(R)}\right]_\S
.\ee
Substituting expressions (\ref{1.e}), we obtain 
\ba 
\l{4.e} 
&{}&{K}_{\tau\tau}^\pm=-\f{ 2\ddot{R} +f^\pm_{,r}(R)} {2\sqrt{f^\pm(R)+\dot{R}^2}},\\
\l{5.e}
&{}&\h{K}^\pm_{\theta \theta } = R\sqrt{f^\pm(R)+\dot{R}^2},\\ 
\l{6.e}
&{}&\h{K}^\pm_{\p \p } = R\sqrt{f^\pm(R)+\dot{R}^2}\sin^2\theta .
\ea 
It is seen that extrinsic curvatures in both  
 parametrizations coincide.

%%%%%%%%%%%%%%%%%%%%%%%%%%%%%%%%%%%%%%%%%%%%%%%%%%%%%%%%%%%%
\section{Israel junction conditions }
%%%%%%%%%%%%%%%%%%%%%%%%%%%%%%%%%%%%%%%%%%%%%%%%%%%%%%%%%%%%%

Next, we consider the Einstein equations and  the Israel junction conditions.
The energy-momentum tensor is taken in the form
\be
\l{21.c}
T_{\m\n}=\th (\eta )S^+_{\m\n} +\th(-\eta )S^-_{\m\n} +\d (\eta )\d_\m^i \d_\n^j S_{ij}
\ee
Because the values of the extrinsic curvatures at the opposite sides of $\S$ are different,
 the derivative
of the extrinsic curvature through the surface $\S$ contains $\d$-singularity.
From the singular part of the $(i j)$ component of the Einstein equations projected on $\S$,
${}^{(3)} R^i{}_j-\d^i_j{}^{(3)} R-
(K^i_j-\d^i_j K)_{,\eta} -KK^i{}_j+(K^2 +K^i_j K^j_i )/2 =8\pi GT^i{}_j$,
follow the relations
\be
\l{51.c}
[K^{i}{}_j ]-\d^i{}_j[ K ]= -8\pi G S^i{}_j ,
\ee
where $[K^{i}{}_j ]=K^{i+}{}_j -K^{i-}{}_j ,\,\,\, K=K^i{}_i$.
The $({\eta}\,{i})$ component of the Einstein
 equations, $-K^{j}{}_{i{}|j}+K_{,i}=
8\pi G T^\eta{}_i $, (vertical bar stands for covariant
derivative with respect to the metric  (\ref{4.a}))
yields
\be
\l{31.c}
K^{j\pm}{}_{i|j}-K^\pm_i =8\pi G \th (\pm\eta )S^{\eta\pm}_i
.\ee
From the  $(\eta\,\eta )$ component of the Einstein equations,
$-{}^{(3)}R/2 -K^i_j K^j_i /2 -K^2 /2 =8\pi GT^\eta_\eta,$
it follows that
\be
\l{41.c}
-\f{1}{2}(K^i_j K^j_i+K^2)^\pm =8\pi G \th (\pm\eta )S^{\eta\pm}{}_\eta
\ee
Further restrictions on $S_{\m\n}^\pm$  follow from the conservation equations of the
energy-momentum tensor \cite{blau}. 

Projections of the components of the bulk metric $g_{\m\n}$ 
 defined as $\t{g}_{ij}= g_{\m\n}\pa x^\m/\pa x^i\, \pa
x^\n/\pa x^j|_{\S}$ are
$$
\t{g}_{\tau\tau}=-1,\qquad \t{g}_{\theta\theta}=R^2 (\tau),
\qquad \t{g}_{\p\p}= R^2 (\tau)\sin^2 \theta .
$$
Projections of the tangent vector $\t{u}_i
= u_\m \pa x^\m/\pa x^i|_{\S}$ are 
\be 
\l{19.b} 
\t{u}_\tau =-1, \qquad
\t{u}_\theta =\t{u}_\p =0 
\ee

Assuming that the energy-momentum tensor $S_{ij}$ has the form
$S_{ij}= \t\s \t{u}_i\t{u}_j +\t\zeta (\t{g}_{ij} -\t{u}_i \t{u}_j)$, 
from  (\ref{51.c}) we obtain
 \be
\l{1.c}
S_\tau^\tau =-\t\s +2\t\zeta,\qquad 
S_\theta^\theta= S_\p^\p =\t\zeta . 
\ee
From the Israel conditions (\ref{31.c}) written as  
$[K^i{}_{j}] = -8\pi G(S^i{}_{j}-\f{1}{2}\d^i_jS)$
we have
\be
[K_\tau^\tau] =4\pi G\t\s ,\qquad
[K_\theta^\theta ] =[K^\p_\p ] =-8\pi G (\t\s/2 -\t\zeta)
\ee 
In the case of the metric (\ref{2.a}) with $f$ independent
of $t$  from (\ref{4.e})- (\ref{6.e}) one obtains 
\ba 
\l{2.c} 
K^{\pm\tau}_\tau=\f{d/d\tau\sqrt{f^\pm (R)+\dot{R}^2}}{\dot{R}}\\
K^{\pm \theta}_\theta = K^{\pm\p}_\p =\f{\sqrt{f^\pm (R)+\dot{R}^2} }{R}.
\ea
Israel conditions take a simple form in the case $\t\zeta =0$ and $\dot{\t\s}=0$. 
From the Israel conditions  it follows that
\ba
\l{3.c}
\f{1}{\dot{R}}\f{d}{d\tau}
\sqrt{{\dot{R}}^2 +f^+(R)}- \sqrt{{\dot{R}}^2 +f^-(R)} =4\pi \t\s,\\
\l{4.c}
\f{1}{R}
\sqrt{{\dot{R}}^2 +f^+(R)}- \sqrt{{\dot{R}}^2 +f^-(R)} = -4\pi \t\s.\\
\ea
Solving the system, we find that $R\t\s =const$ and
\be
\l{5.c}
R\left(\sqrt{{\dot{R}}^2 +f^+(R)}- \sqrt{{\dot{R}}^2 +f^-(R)}\right)=M=const
\ee
Relation (\ref{5.c}) can be rewritten as (cf.\cite{chowdh})
\be
\l{jc}
{\dot{R}}^2  +f^\pm (R) -\f{1}{M^2} \left[ m_+ -m_- \mp\f{M^2}{2R}\right]^2  =0.
\ee

\section{Conclusions}

Geometry of two useful coordinate systems used to study dynamic of thin shells, 
Kodama-Schwarzschild coordinates and Gaussian normal coordinate system, was studied.
 Transformation between
the coordinate systems is considered and explicitly constructed for the case of 
Kodama-Schwarzschild metric independent of time.
Extrinsic curvatures of the surface swept by the shell are calculated  
for a general time-dependent metric  in both Kodama-Schwarzschild
and normal Gaussian parametrizations and are shown to give the same result.
Application to the Israel junction conditions is discussed.

\section{Acknowledgments}

The work was supported by Skobeltsyn Institute of Nuclear Physics.

\end{document}